\documentclass[aip,adv,reprint,nolongbibliography]{revtex4-2}
\usepackage{graphicx}
\usepackage{bm}
\usepackage{amsmath}
\usepackage{color}

\begin{document}


\title{Spin-1/2 Heisenberg diamond-like decorated honeycomb lattice in a magnetic field from the perspective of localized magnons and exact diagonalization}
\author{Katar\'ina Kar{l}'ov\'a}
\email{katarina.karlova@upjs.sk}
\affiliation{Faculty of Science, P.~J.\ \v{S}af\'{a}rik University, Park Angelinum 9, Ko\v{s}ice, Slovakia}
\author{Malo Rouxel}
\affiliation{\scalebox{0.98}[1] {Laboratoire de Physique Th\'eorique et
Mod\'elisation, CNRS UMR 8089, CY Cergy Paris Universit\'e, France}}
\author{Jozef Stre\v{c}ka}
\affiliation{Faculty of Science, P.~J.\ \v{S}af\'{a}rik University, Park Angelinum 9, Ko\v{s}ice, Slovakia}
\author{Andreas Honecker}
\affiliation{\scalebox{0.98}[1] {Laboratoire de Physique Th\'eorique et
Mod\'elisation, CNRS UMR 8089, CY Cergy Paris Universit\'e, France}}

\date{\today}
	
\begin{abstract}
The spin-$1/2$ quantum Heisenberg model on the diamond-decorated honeycomb lattice in a magnetic field is investigated using complementary
analytical and numerical methods. The ground-state phase diagram comprises Lieb--Mattis ferrimagnetic, dimer--tetramer, monomer--dimer, spin-canted, and saturated paramagnetic phases, giving rise to characteristic magnetization plateaus, continuous and discontinuous field-induced transitions. In the highly frustrated regime, a localized-magnon approach based on an effective classical lattice-gas model provides an accurate description of the low-temperature magnetism and thermodynamics in agreement with numerical results. At intermediate frustration, finite-temperature calculations reveal characteristic specific-heat signatures of the field-induced transitions including the development of Ising critical behavior at the field-induced transition between the monomer--dimer and Lieb--Mattis ferrimagnetic phase.
\end{abstract}
	
\maketitle
\onecolumngrid
	
\section{Introduction}\label{Sec:introduction}

Geometrically frustrated quantum spin systems on decorated lattices provide a natural setting for unconventional ground states including macroscopically degenerate phases, magnetization plateaus, and remarkable field-induced phase transitions.\cite{Shibata2016,Hirose2016,Hirose2018,Caci2023,Karlova2024,Dmitriev2025,KarlovaZNA2026} In highly frustrated regimes, exact localized-magnon states may further allow the relevant low-energy degrees of freedom to be mapped onto effective classical lattice-gas models, providing a simple analytical description of low-temperature thermodynamics.\cite{Derzhko2007} Diamond-decorated lattices constitute a particularly rich example, displaying Lieb--Mattis ferrimagnetic, dimer--tetramer, monomer--dimer, and spin-canted phases, as well as field-induced critical behavior.\cite{Shibata2016,Hirose2016,Hirose2018,Caci2023,Karlova2024,Dmitriev2025,KarlovaZNA2026} Recently, a weak distortion of the diamond-decorated honeycomb lattice was shown to generate effective dimer physics and signatures of a Kasteleyn transition.\cite{Karlova2026}

Here, we investigate the undistorted spin-$1/2$ Heisenberg model on the diamond-decorated honeycomb lattice in a magnetic field. We determine its ground-state phase diagram and investigate its finite-temperature magnetic properties. Particular attention is paid to the validity of the effective lattice-gas description in the highly frustrated regime and to finite-temperature signatures of the field-induced transitions.

\section{MODEL AND METHOD}
\label{model}
\begin{figure}[t]
    \centering
    \includegraphics[width=0.25\columnwidth]{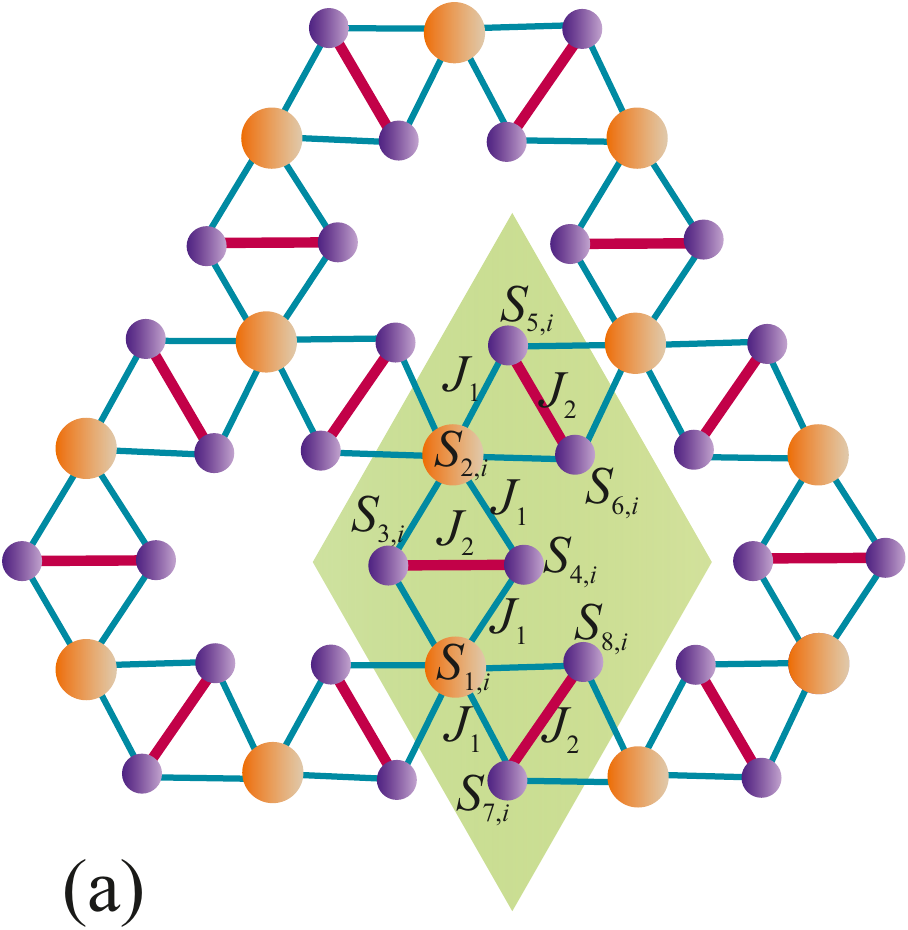}
    \hspace{2cm}
    \includegraphics[width=0.35\columnwidth]{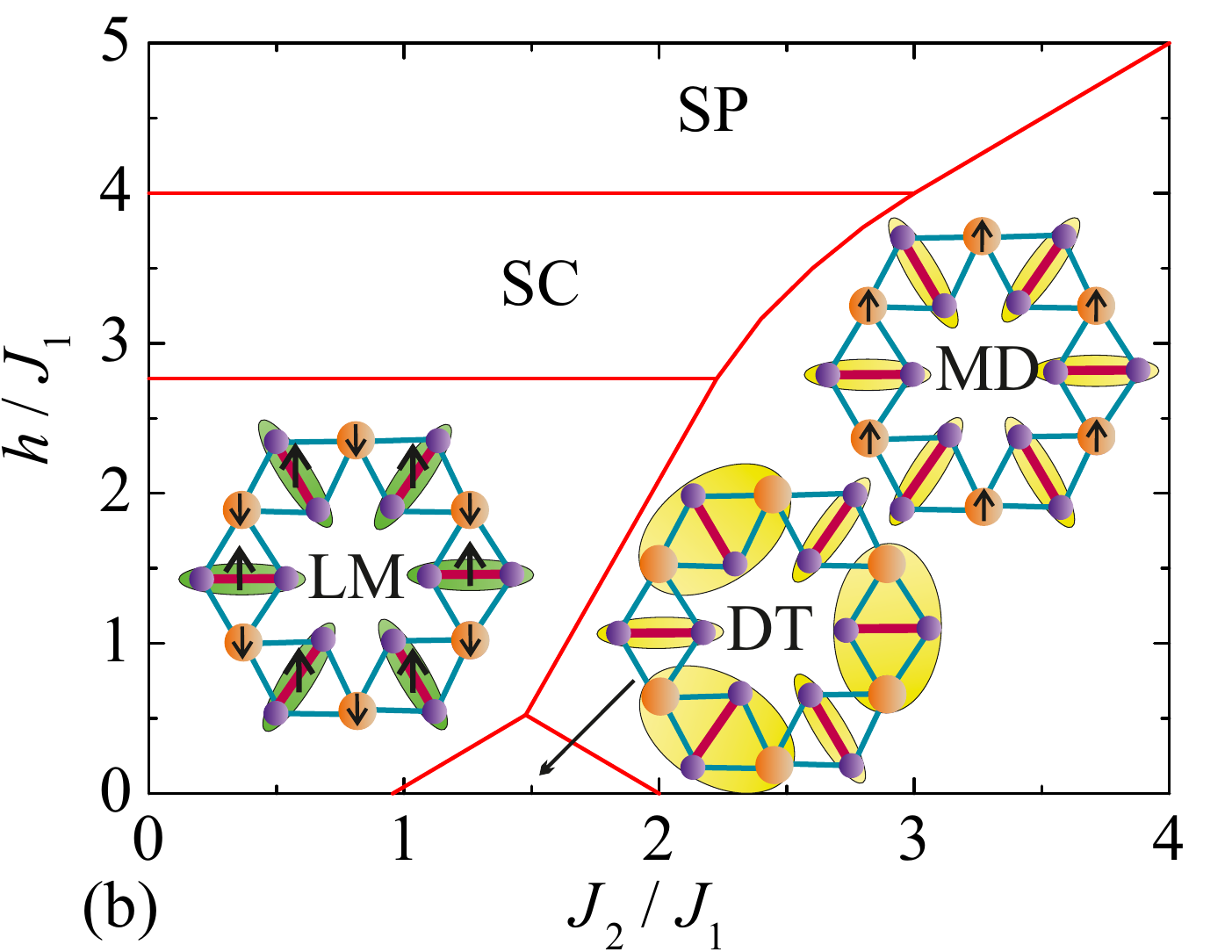}
		\vspace{-0.5cm}
  \caption{(a) Schematic illustration of the spin-$1/2$ Heisenberg model on a diamond-decorated honeycomb lattice with the green rhombus indicating the unit cell. (b) Ground-state phase diagram in the $J_2/J_1$--$h/J_1$ parameter plane, where small yellow (green) ellipses denote singlet (triplet) dimers and large yellow ellipses denote singlet tetramers.}
    \label{fig:model_gspd}
\end{figure}

We consider the spin-$1/2$ Heisenberg model on a diamond-decorated honeycomb lattice in an external magnetic field, which is schematically illustrated in Fig.~\ref{fig:model_gspd}(a) and defined by the Hamiltonian
{
\setlength{\abovedisplayskip}{2pt}
\setlength{\belowdisplayskip}{0pt}
\setlength{\abovedisplayshortskip}{2pt}
\setlength{\belowdisplayshortskip}{0pt}
\setlength{\jot}{0pt}
\begin{equation}
\begin{split}
\hat{\mathcal H} = \sum_i \Big\{ &
J_1 \Big[
(\hat{\bm{S}}_{1,i}+\hat{\bm{S}}_{2,i})
\cdot(\hat{\bm{S}}_{3,i}+\hat{\bm{S}}_{4,i})
+(\hat{\bm{S}}_{5,i}+\hat{\bm{S}}_{6,i})
\cdot(\hat{\bm{S}}_{2,i}+\hat{\bm{S}}_{1,k})
+(\hat{\bm{S}}_{7,i}+\hat{\bm{S}}_{8,i})
\cdot(\hat{\bm{S}}_{1,i}+\hat{\bm{S}}_{2,l})\Big]
\\[-10pt]
&+J_2(\hat{\bm{S}}_{3,i}\cdot\hat{\bm{S}}_{4,i}
+\hat{\bm{S}}_{5,i}\cdot\hat{\bm{S}}_{6,i}
+\hat{\bm{S}}_{7,i}\cdot\hat{\bm{S}}_{8,i})
-h\sum_{j=1}^{8}\hat S^z_{j,i}\Big\}.
\end{split}
\label{hamos}
\end{equation}
}%
Here, $\hat{\bm{S}}_{j,i}$ denotes a spin-$1/2$ operator at the $j$th site of the $i$th unit cell, the indices $k$ and $l$ denote adjacent unit cells. The antiferromagnetic coupling $J_1>0$ describes the interaction between the nearest-neighbor nodal and decorating spins, whereas the antiferromagnetic interaction $J_2>0$ couples nearest-neighbor decorating spins forming the dimers. The last term accounts for the Zeeman interaction with an external magnetic field $h$ applied along the $z$ direction.

The ground-state phase diagram was determined using the density-matrix renormalization group (DMRG) method \cite{White1992} implemented within the ALPS project.\cite{Bauer2011} We consider $L\times L$ lattices with periodic boundary conditions, where $L$ is the linear lattice size and $N=8L^2$ denotes the total number of spins. Zero-temperature magnetization curves were obtained by Lanczos exact diagonalization (EDL) \cite{Lanczos1950} for $L=2$ and DMRG \cite{White1992} for $L=4$. Finite-temperature properties were calculated using quantum Monte Carlo (QMC) simulations in the dimer basis \cite{Weber2022,Caci2023} for $L=6$ and a combination of full exact diagonalization and the finite-temperature Lanczos method (FTLM) \cite{prelovsek_bonca_2013,Wietek2019} for $L=2$. The latter exploits the local conservation of the total spin of each decorating dimer and follows the approach of Ref.~\onlinecite{Karlova2026}. For all finite-temperature calculations, we set $k_{\rm B}=1$.

In the highly frustrated regime, the low-energy states can be described within the localized-magnon theory by mapping the singlet of the decorating dimers onto a quasiparticle of an effective lattice-gas model. We follow the approach developed for the diamond-decorated square lattice in Ref.~\onlinecite{Caci2023} by adapting it to the underlying honeycomb-lattice geometry. The occupation number $n_i=0,1$
of a quasiparticle corresponds to the triplet and singlet states of a decorating dimer, respectively. In the present case, the Hamiltonian of the effective lattice-gas model accounting for a single species of localized magnons with the chemical potential $\mu=J_1+J_2-h$ is given by
{
\setlength{\abovedisplayskip}{1pt}
\setlength{\belowdisplayskip}{0pt}
\setlength{\abovedisplayshortskip}{0pt}
\setlength{\belowdisplayshortskip}{0pt}
\begin{equation}
\mathcal{H}_{\rm eff}
=E_{\rm FM}-\mu\sum_i n_i-h\sum_j (S_{1,j}^z + S_{2,j}^z)
\label{eq:Heff}
\end{equation}
}%
when expressed relative to the fully polarized state with the energy $E_{\rm FM}$.
Here, the first summation runs over all bonds (dimers) of the underlying lattice and the last term incorporating a summation over all unit cells represents the Zeeman energy of the nodal spins. The resulting partition function can be evaluated exactly in the thermodynamic limit, from which all quantities can be 
derived (see  Ref.~\onlinecite{Caci2023}).

\section{Results and Discussion}\label{Results}
The ground-state phase diagram shown in Fig.~\ref{fig:model_gspd}(b) comprises five phases: the Lieb--Mattis ferrimagnetic (LM), macroscopically degenerate dimer--tetramer (DT), monomer--dimer (MD), spin--canted (SC), and saturated paramagnetic (SP) states, whose representative configurations are also shown there. The LM phase consists of triplet dimers with the nodal spins aligned preferentially antiparallel to them. In the DT phase, singlet dimers and tetramers can be distributed in macroscopically many ways, subject to the local constraint that each singlet tetramer is surrounded by singlet dimers.\cite{Shibata2016} In the MD phase, all decorating dimers form singlets, while the nodal spins are free at zero field and become fully polarized in a finite field. At lower frustration and sufficiently strong fields, the LM state evolves through the gapless SC phase with continuously varying magnetization into the saturated paramagnet.

\begin{figure}[ht!]
	\centering
	\includegraphics[width=0.3\textwidth]{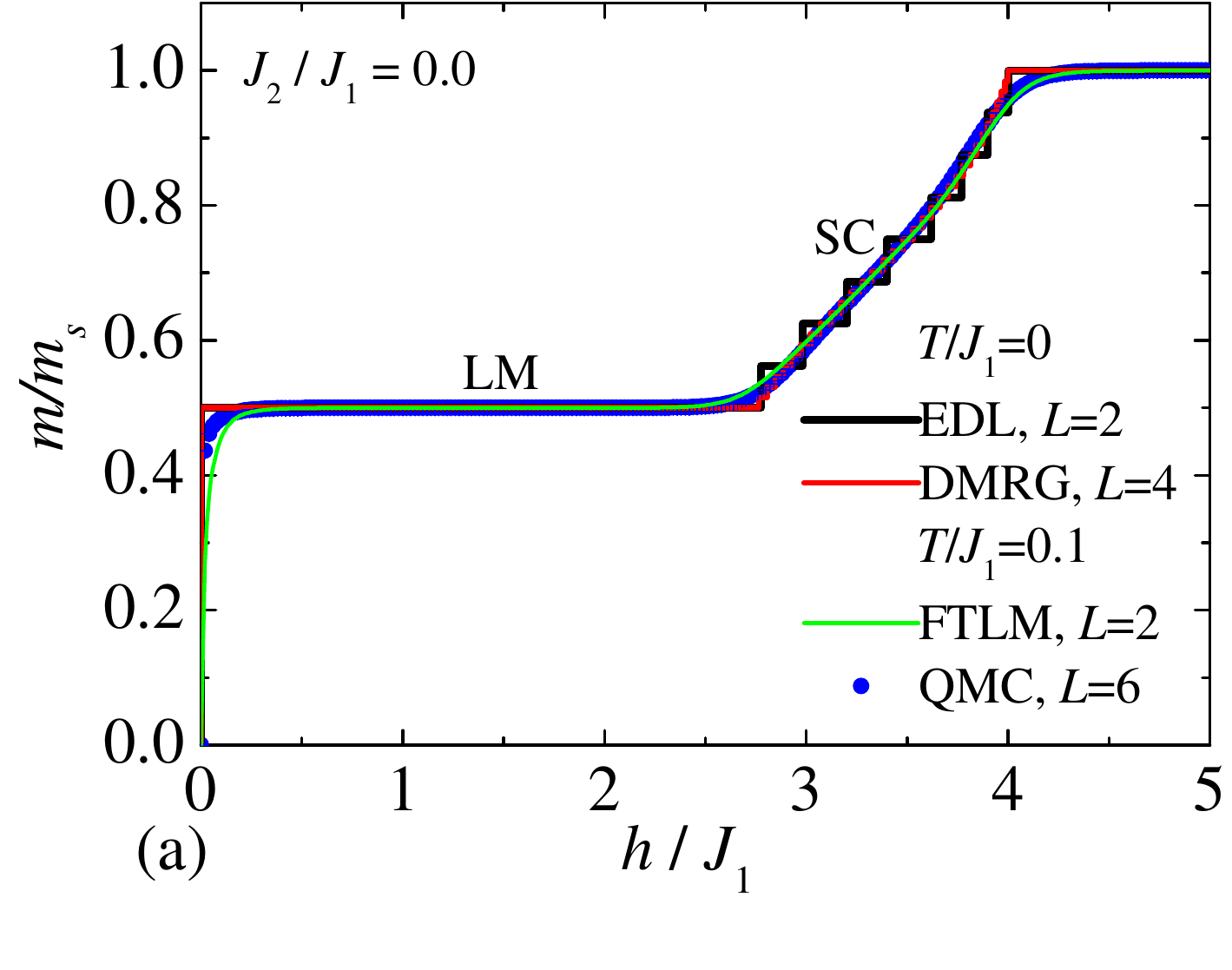}
	\includegraphics[width=0.3\textwidth]{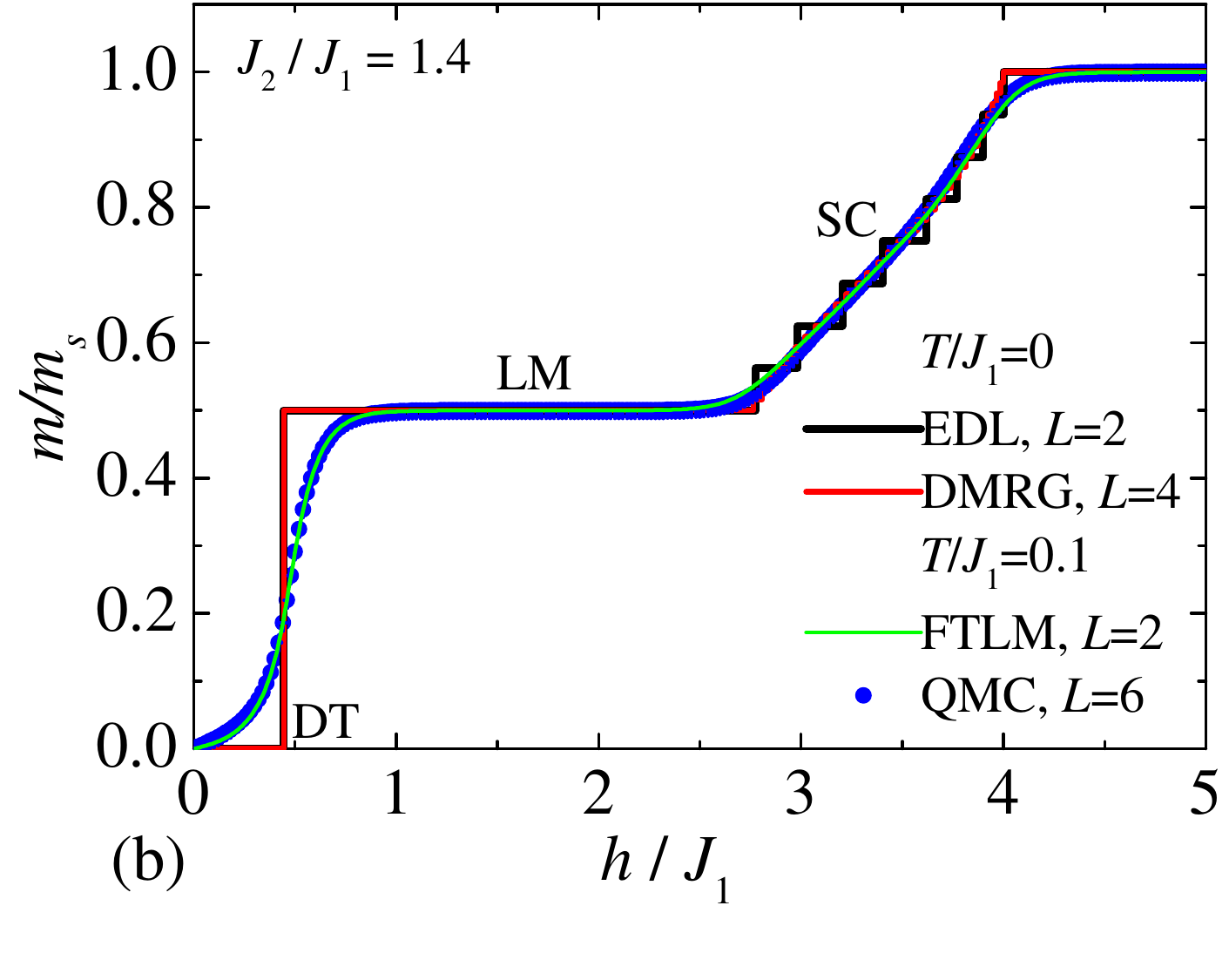}
\includegraphics[width=0.3\textwidth]{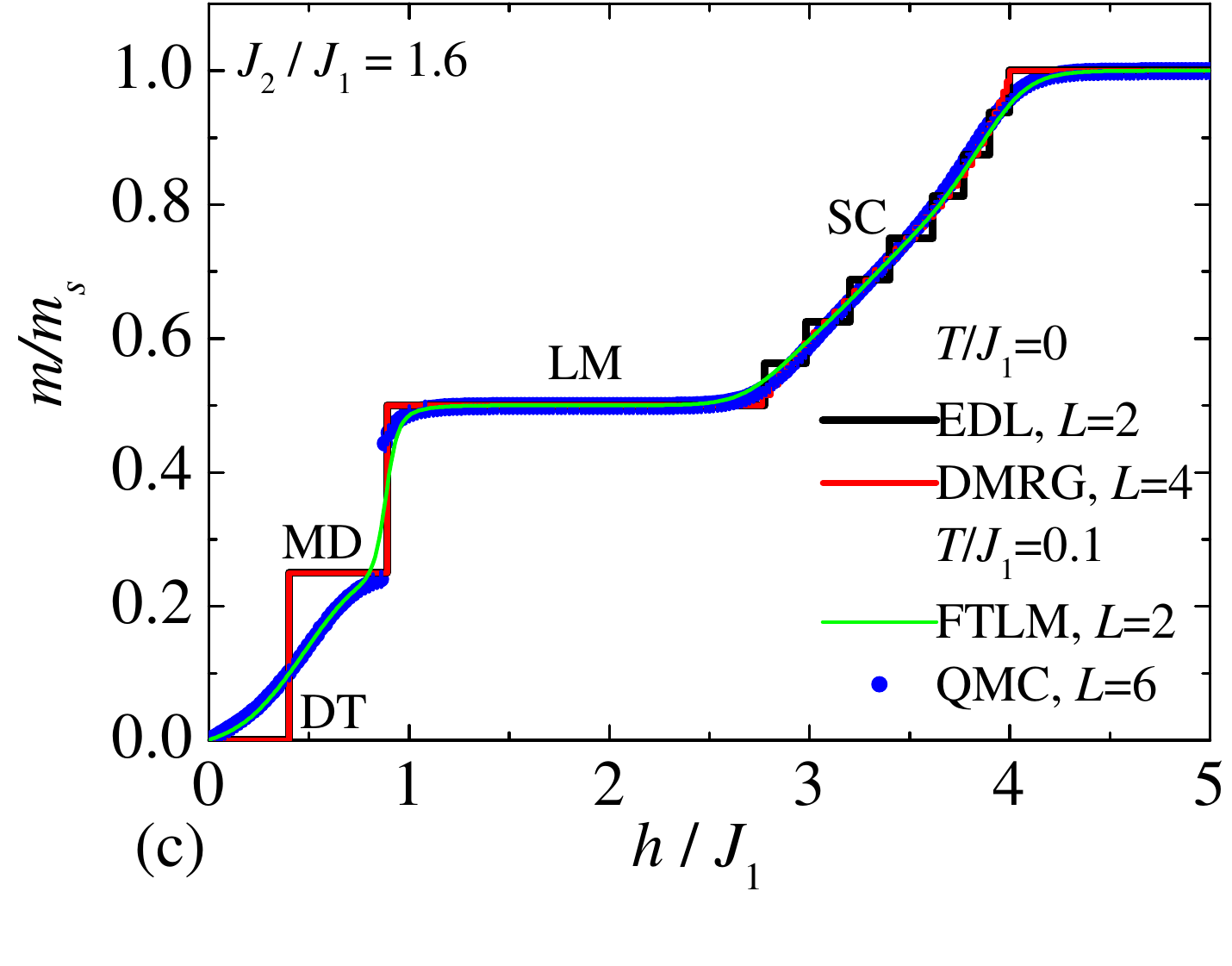} 
\raisebox{-3mm}{\includegraphics[width=0.3\textwidth]{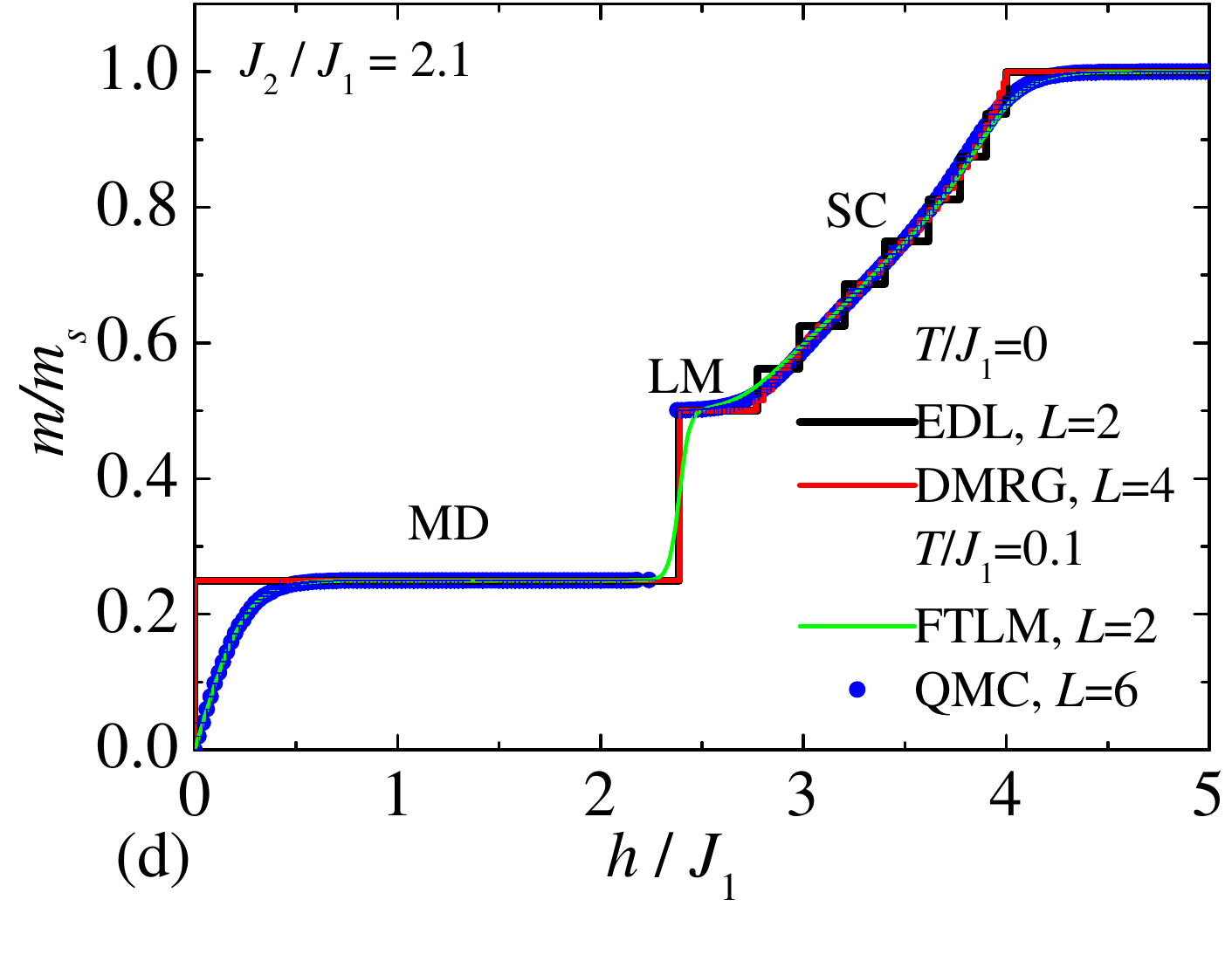}}
\includegraphics[width=0.3\textwidth]{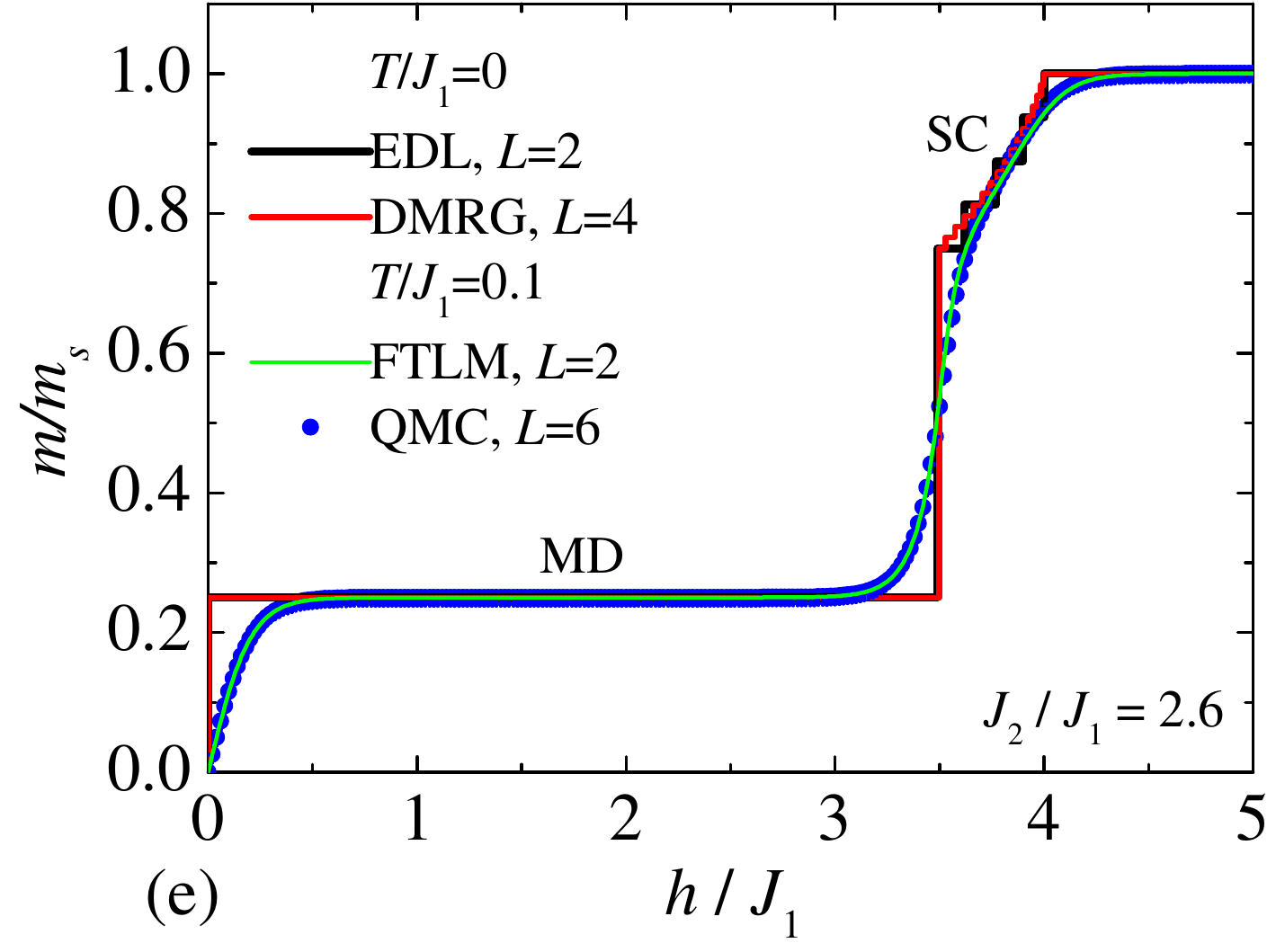}
\includegraphics[width=0.3\textwidth]{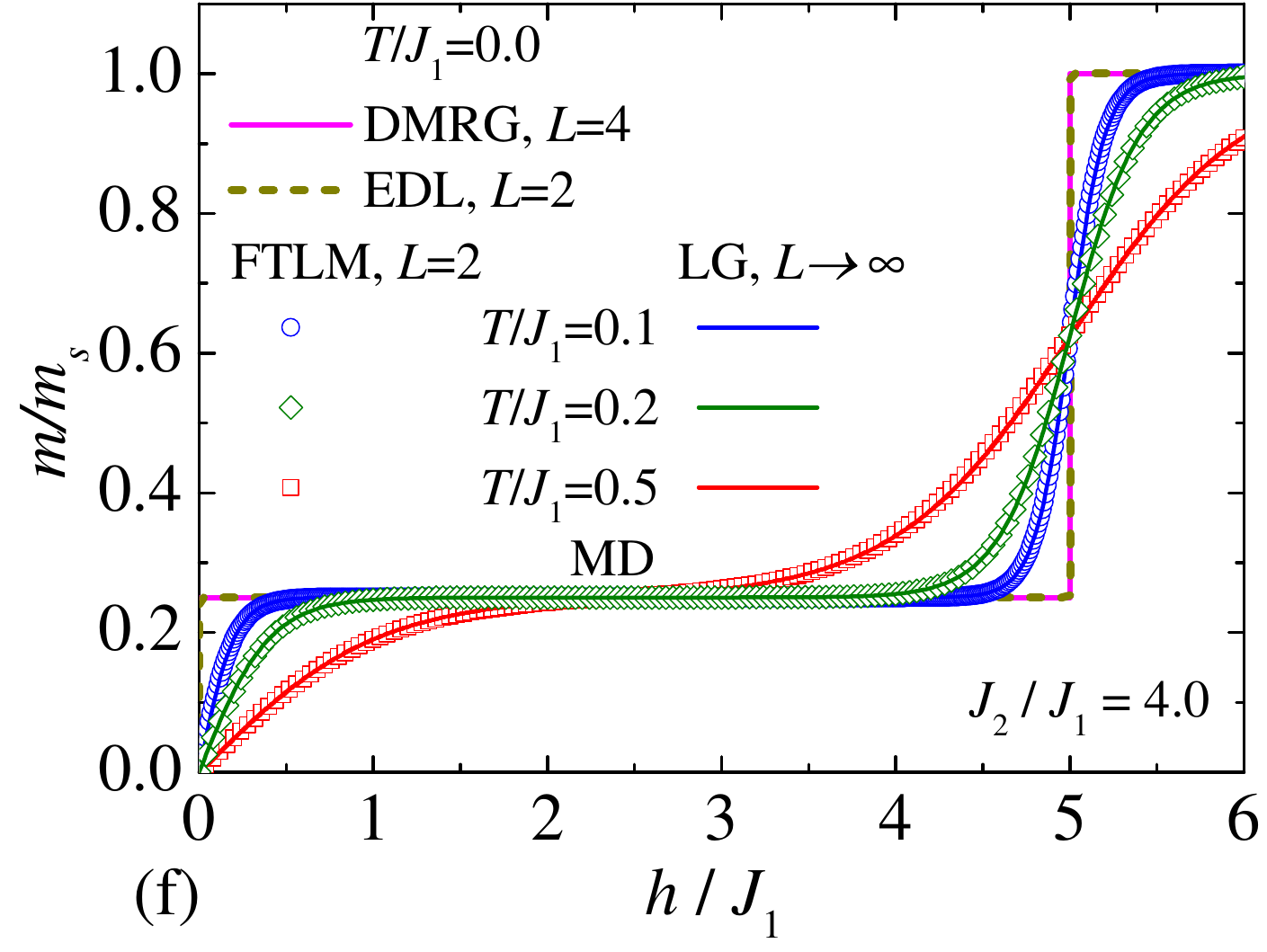}
	\vspace{-0.6cm}
	\caption{Representative isothermal magnetization curves calculated for six selected values of the interaction ratio $J_2/J_1=0.0$, $1.4$, $1.6$, $2.1$, $2.6$, and $4.0$. Zero-temperature results obtained by DMRG ($L=4$) and EDL ($L=2$) are compared with finite-temperature QMC ($L=6$) and FTLM ($L=2$) results at $T/J_1=0.1$. For $J_2/J_1=4.0$, the FTLM results are additionally compared at several temperatures with analytical results derived from the effective lattice-gas model.}
	\label{fig:mag}
\end{figure}
Figure~\ref{fig:mag} shows representative isothermal magnetization curves for a few selected values of the interaction ratio $J_2/J_1$ corresponding to typical vertical cuts through the ground-state phase diagram in Fig.~\ref{fig:model_gspd}(b). For $J_2/J_1=0$, increasing the field drives the system from the LM phase through the SC phase into the SP. At $J_2/J_1=1.4$, the process starts from the DT phase, while at $J_2/J_1=1.6$ a narrow MD phase appears between the DT and LM phases. The cuts at $J_2/J_1=2.1$ and $2.6$ show the MD--LM--SC and MD--SC sequences, respectively, whereas at $J_2/J_1=4$ the MD state undergoes a direct field-induced transition toward the SP. The DT, MD, and LM phases correspond to zero, one-quarter, and one-half magnetization plateaus, respectively. At $J_2/J_1=4$, the effective lattice-gas model shows excellent agreement with the FTLM results over the investigated temperature range.

The zero-temperature DMRG and EDL results for different system sizes are consistent with each other showing weak finite-size effects within the gapped phases and sizable finite-size effects within the gapless SC phase. The discrete magnetization steps within the SC phase are thus expected to evolve into a continuous curve in the thermodynamic limit. At $T/J_1=0.1$, FTLM and QMC consistently show thermal rounding of the field-induced transitions and magnetization plateaus. Their agreement for $L=2$ and $6$ indicates weak finite-size effects at this temperature. For $J_2/J_1=4$, the progressive thermal smearing of the magnetization plateau is further illustrated at several temperatures.

\begin{figure}[t!]
	\centering
	\includegraphics[width=0.3\textwidth]{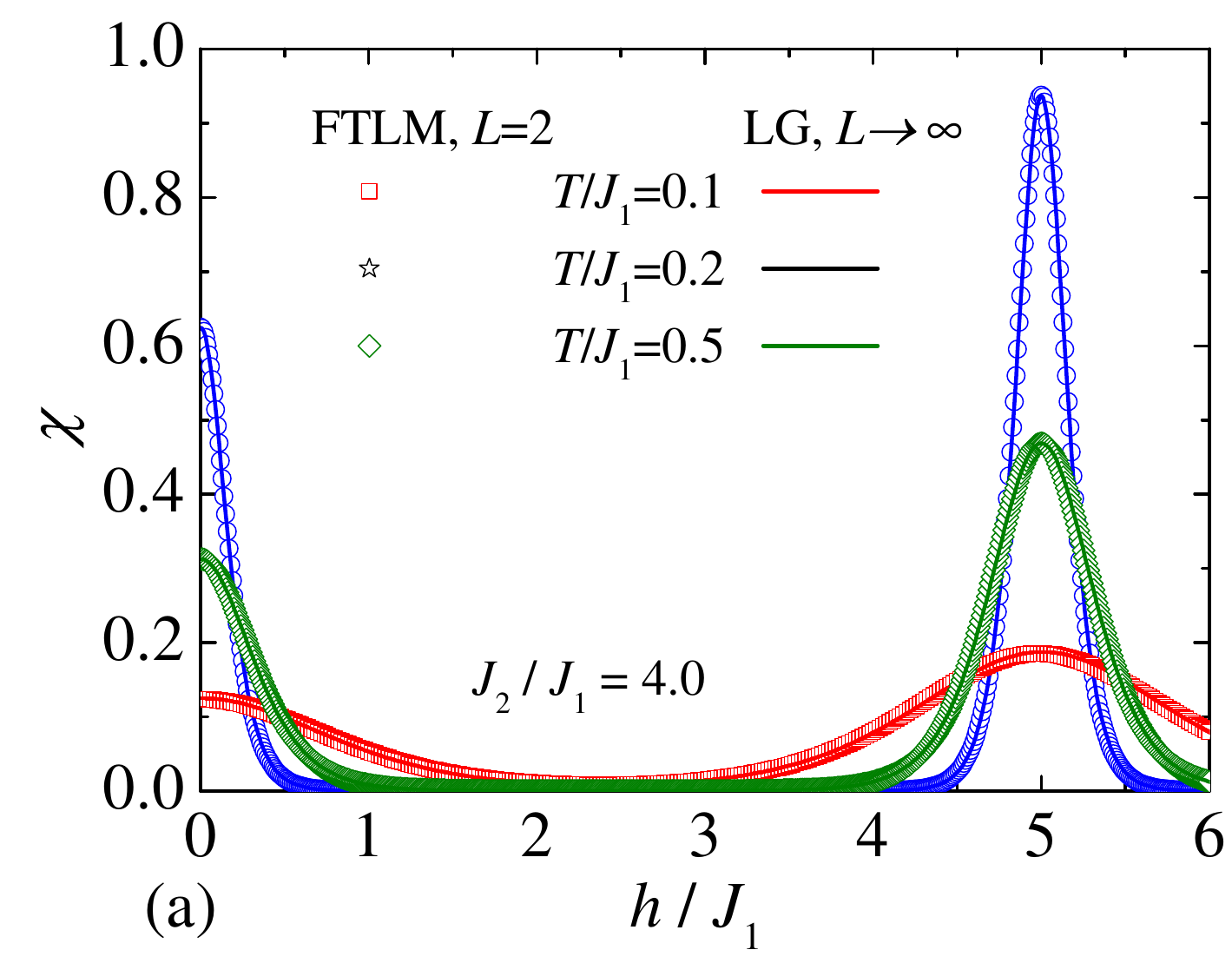}
	\includegraphics[width=0.3\textwidth]{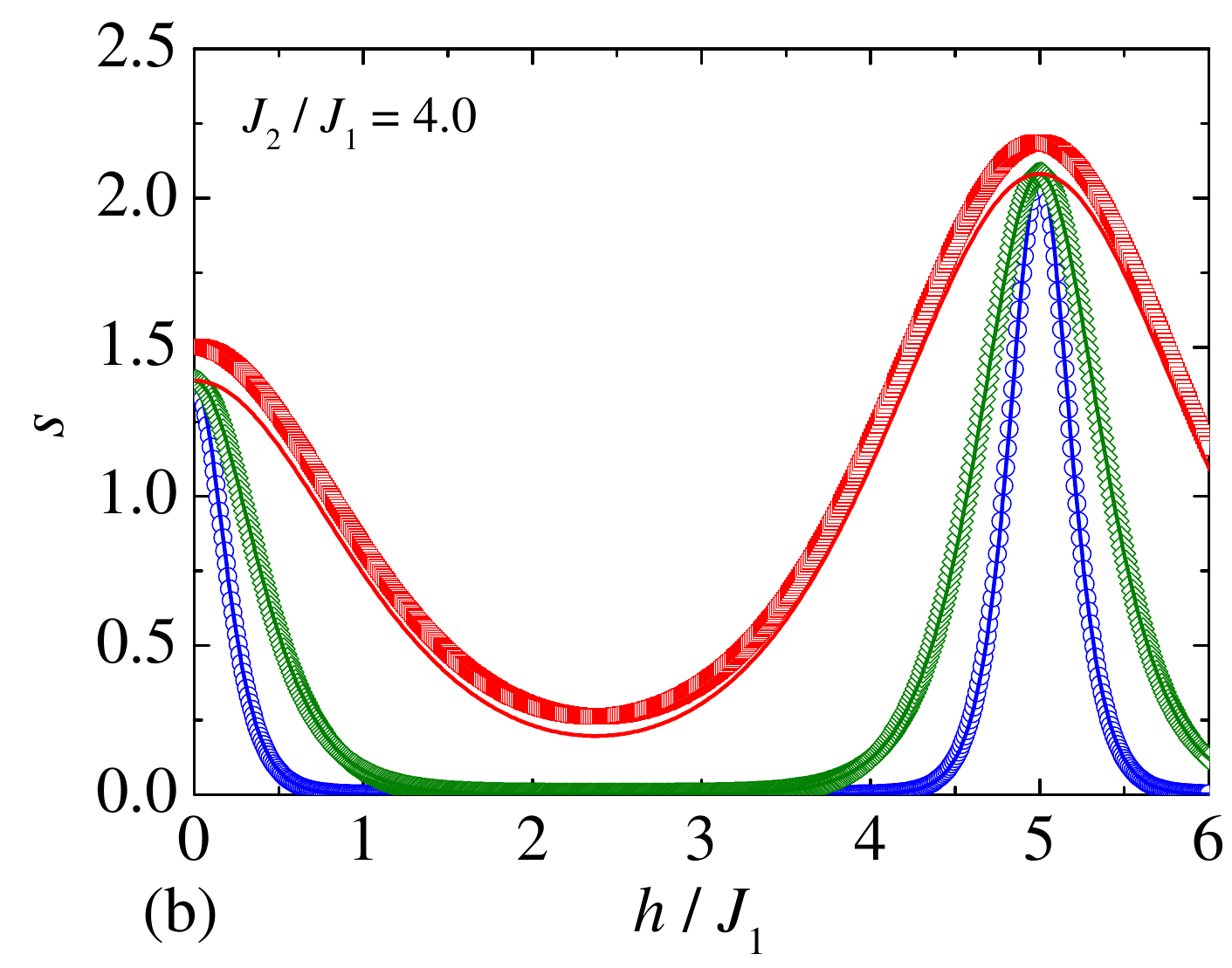}
	\includegraphics[width=0.3\textwidth]{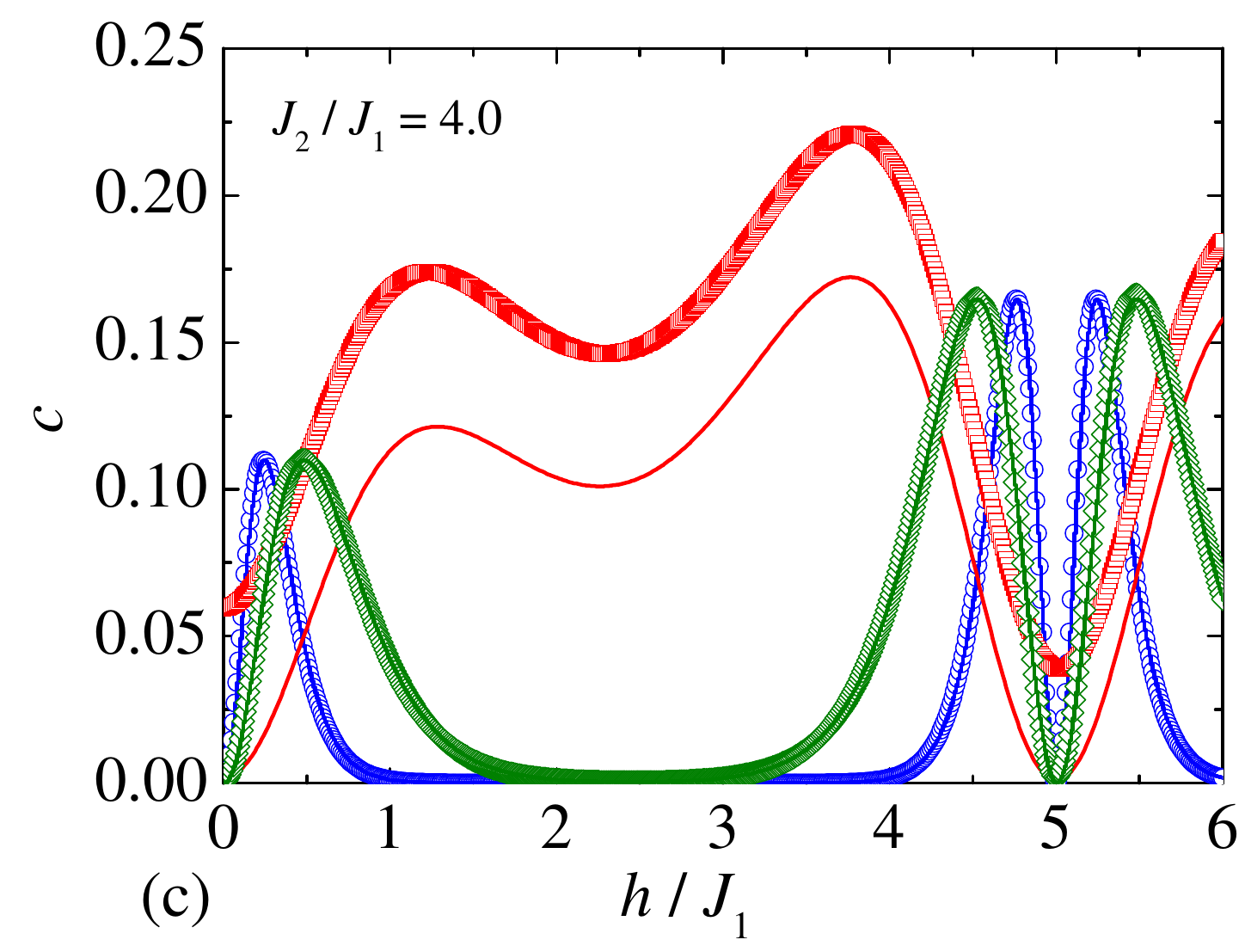}
	\vspace{-0.5cm}
	\caption{(a) Magnetic susceptibility $\chi$ per spin, (b) entropy per unit cell $s$, and (c) specific heat $c$ per spin as a function of the magnetic field $h/J_1$ for $J_2/J_1=4.0$ at three selected temperatures. Symbols show the FTLM results for $L=2$, while solid lines denote the effective lattice-gas (LG) results in the thermodynamic limit. The legend in (a) applies to all panels.}
	\label{fig:thermo}
\end{figure}
Figure~\ref{fig:thermo} compares the FTLM results for 32 spins with the thermodynamic-limit lattice-gas description at $J_2/J_1=4$. The most intriguing features occur at the transition fields $h/J_1=0$ and $5$. The susceptibility exhibits a single peak at each transition progressively suppressed with increasing temperature, while the entropy behaves similarly with the maxima broadening and gradually increasing with temperature. The specific heat displays a characteristic two-peak structure around $h/J_1=5$, whereas only a single peak is observed slightly above zero field. The lattice-gas model reproduces the susceptibility and entropy remarkably well up to $T/J_1\lesssim 0.5$, while the specific heat is more sensitive to thermal excitations beyond the effective lattice gas and is captured only qualitatively at the highest temperature. The close agreement between the finite-size FTLM and thermodynamic-limit results reflects the fragmented nature of the MD phase and the simple fully polarized high-field state resulting in negligible finite-size effects.

\begin{figure}[t!]
	\centering
	\includegraphics[width=0.3\textwidth]{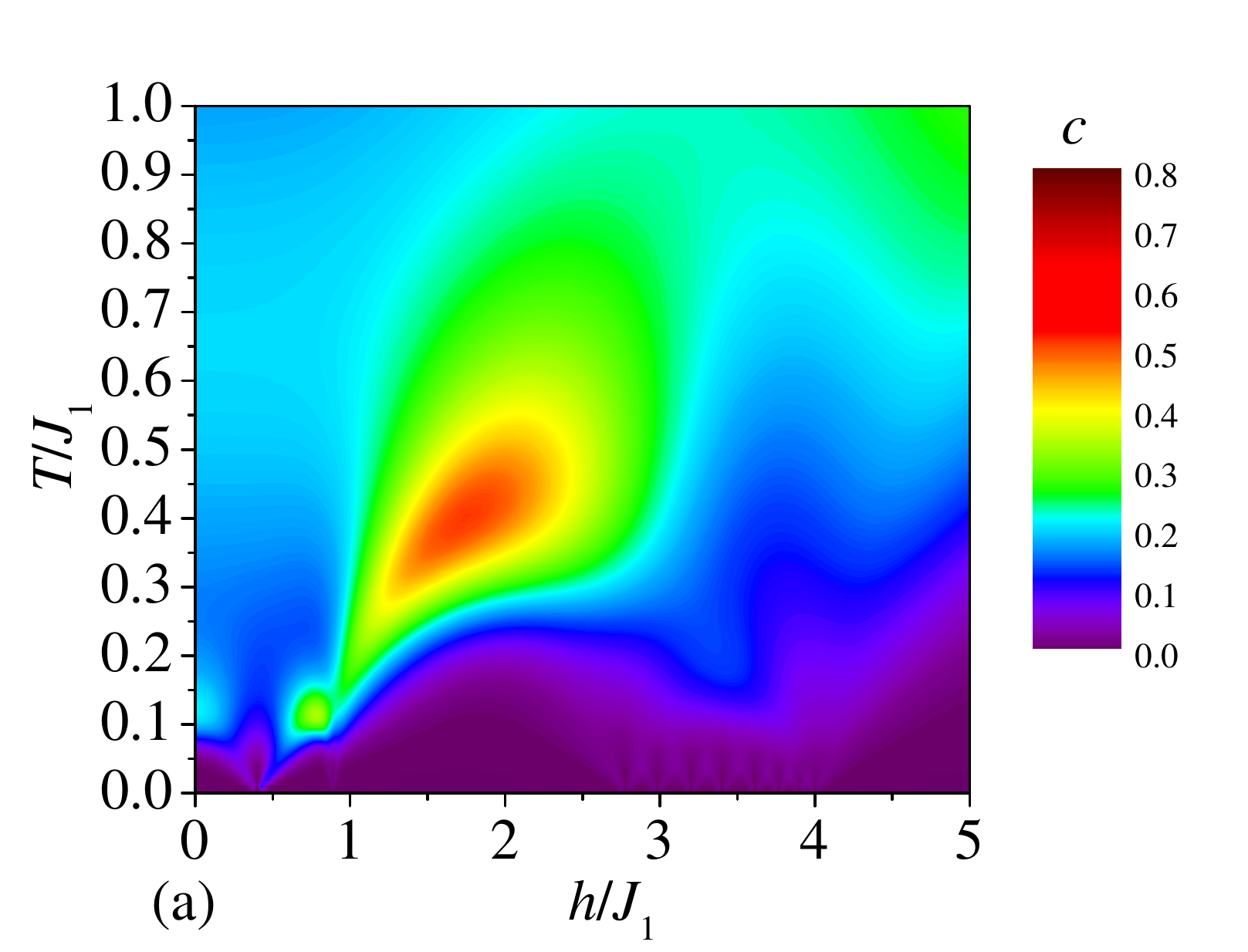}
	\includegraphics[width=0.3\textwidth]{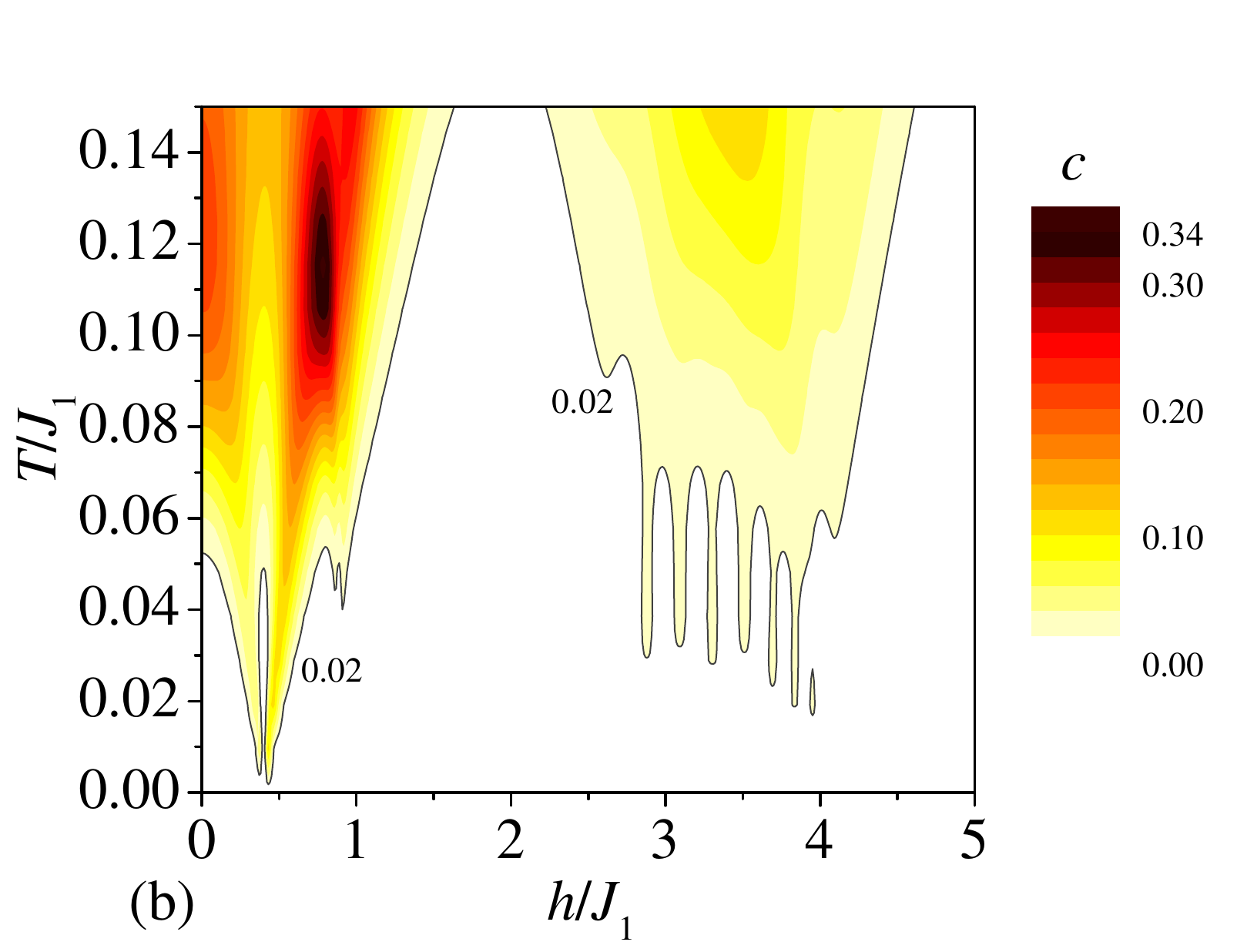}
	\includegraphics[width=0.3\textwidth]{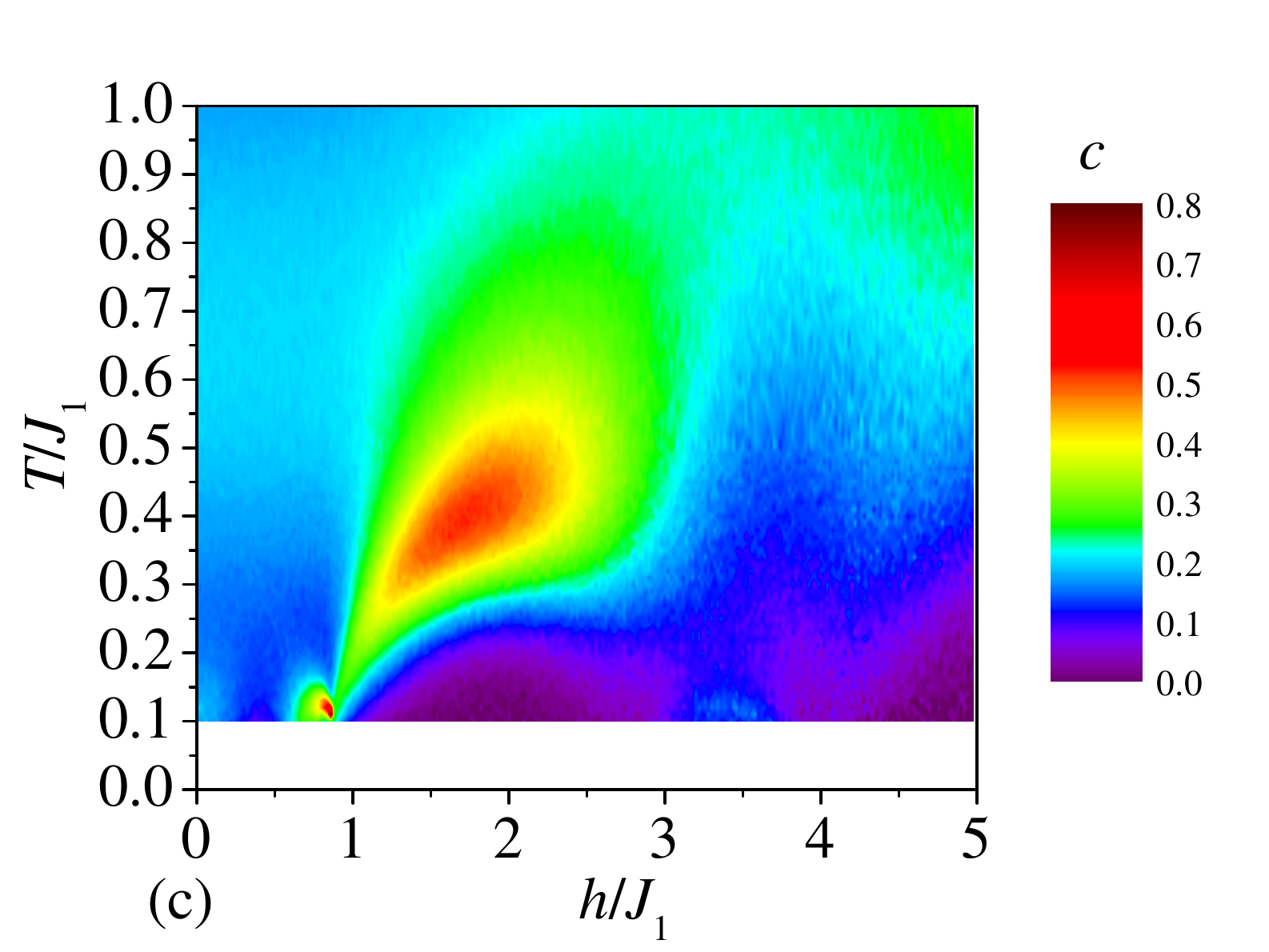}
	\vspace{-0.5cm}
	\caption{Specific heat $c$ as a function of the magnetic field $h/J_1$ and temperature $T/J_1$ for $J_2/J_1=1.6$.
(a) FTLM results for $L=2$, (b) low-temperature zoom of panel (a), and (c) QMC results for $L=6$. Panel (b) uses a different color scale for clarity, with the gray contour marking $c=0.02$.}
	\label{fig:FTLM}
\end{figure}
Figure~\ref{fig:FTLM} shows a density plot of the specific heat in the field-temperature plane for $J_2/J_1=1.6$. FTLM results for $L=2$ in Fig.~\ref{fig:FTLM}(a) are dominated by a broad maximum around $h/J_1\simeq2$ originating from
thermal excitations from the LM ground state. The low-temperature zoom in Fig.~\ref{fig:FTLM}(b) reveals additional features associated with the field-induced transitions. The narrow structures within the SC phase at $3\lesssim h/J_1\lesssim4$ are finite-size effects of the $L=2$ cluster arising from successive crossings between adjacent total-spin sectors. These crossings also produce the EDL magnetization steps in Fig.~\ref{fig:mag}.

At low fields, the enhanced specific heat around $h/J_1\simeq0.4$ is associated with the DT--MD crossover consistent with the low-energy excitation spectrum of the DT phase discussed in Ref.~\onlinecite{Karlova2026}. A second sizable low-temperature peak around $h/J_1\simeq0.9$ relates to the MD--LM thermal transition. Unlike the DT--MD crossover, it is associated with a finite-temperature critical point from the two-dimensional Ising universality class, as previously found for the diamond-decorated square lattice.\cite{Caci2023} Pronounced finite-size effects are therefore expected near this transition. While the FTLM cluster $L=2$ shows only a weak specific-heat maximum around $T/J_1\simeq0.12$, QMC data for $L=6$ in Fig.~\ref{fig:FTLM}(c) indeed exhibit a substantially enhanced response, reflecting the development of Ising critical behavior.

\section{Conclusion}\label{Sec:conclusion}
We have investigated the ground-state and finite-temperature properties of the spin-$1/2$ Heisenberg model on the diamond-decorated honeycomb lattice in a magnetic field. The ground-state phase diagram comprises Lieb--Mattis, dimer--tetramer, monomer--dimer, spin-canted, and saturated paramagnetic phases, giving rise to characteristic magnetization plateaus and field-induced transitions. In the highly frustrated regime, the low-temperature magnetism and thermodynamics are accurately described by an effective classical lattice-gas model derived within the localized-magnon framework. At intermediate frustration, the specific heat reveals additional low-temperature signatures of the field-induced transitions, including the development of Ising critical behavior at the MD--LM transition. The agreement between complementary analytical and numerical approaches provides a consistent description of the model from its ground-state phase diagram to its low-temperature thermodynamics. 

\section*{DATA AVAILABILITY}
The data that support findings of this study are available from the corresponding author upon reasonable request.

\begin{acknowledgments}
We acknowledge support from the Štefánik Slovak–French bilateral program (SK-FR-24-0005/53633XM), the French National Research Agency (ANR-22-CE30-0042), and the Slovak Research and Development Agency (APVV-24-0091). KK thanks S.\ Wessel and N.\ Çaçi for providing the QMC code developed by L.\ Weber and for helpful discussions.
\end{acknowledgments}


\begin{thebibliography}{99}

\bibitem{Shibata2016}
K. Morita and N. Shibata,
J. Phys. Soc. Jpn. \textbf{85}, 033705 (2016).

\bibitem{Hirose2016}
Y. Hirose, A. Oguchi, and Y. Fukumoto,
J. Phys. Soc. Jpn. \textbf{85}, 094002 (2016).

\bibitem{Hirose2018}
Y. Hirose, S. Miura, C. Yasuda, and Y. Fukumoto,
AIP Adv. \textbf{8}, 101427 (2018).

\bibitem{Caci2023}
N. Caci, K. Karl'ov\'a, T. Verkholyak, J. Stre\v{c}ka, S. Wessel,
and A. Honecker,
Phys. Rev. B \textbf{107}, 115143 (2023).

\bibitem{Karlova2024}
K. Karl'ov\'a, A. Honecker, N. Caci, S. Wessel, J. Stre\v{c}ka,
and T. Verkholyak,
Phys. Rev. B \textbf{110}, 214429 (2024).

\bibitem{Dmitriev2025}
D. V. Dmitriev, V. Ya. Krivnov, and O. A. Vasilyev,
Phys. Rev. B \textbf{112}, 094426 (2025).


\bibitem{KarlovaZNA2026}
K. Karl'ov\'a and J. Stre\v{c}ka,
Z. Naturforsch. A \textbf{81}, 319 (2026).



\bibitem{Derzhko2007}
O. Derzhko, J. Richter, A. Honecker, and H.-J. Schmidt,
Low Temp. Phys. \textbf{33}, 745 (2007).

\bibitem{Karlova2026}
K. Karl'ov\'a, A. Rufino, T. Verkholyak, N. Caci, S. Wessel,
J. Stre\v{c}ka, F. Mila, and A. Honecker,
Phys. Rev. Lett. \textbf{137}, 116703 (2026).

\bibitem{White1992}
S. R. White,
Phys. Rev. Lett. \textbf{69}, 2863 (1992).

\bibitem{Bauer2011}
B. Bauer, L. D. Carr, H. G. Evertz, \textit{et al.},
J. Stat. Mech.: Theory Exp. \textbf{2011}, P05001 (2011).

\bibitem{Lanczos1950}
C. Lanczos,
J. Res. Natl. Bur. Stand. \textbf{45}, 255 (1950).

\bibitem{Weber2022}
L. Weber, A. Honecker, B. Normand, P. Corboz, F. Mila, and S. Wessel,
SciPost Phys. \textbf{12}, 054 (2022).

\bibitem{prelovsek_bonca_2013}
P. Prelov\v{s}ek and J. Bon\v{c}a,
in \textit{Strongly Correlated Systems: Numerical Methods},
edited by A. Avella and F. Mancini
(Springer Berlin Heidelberg, 2013), pp. 1--30.

\bibitem{Wietek2019}
A. Wietek, P. Corboz, S. Wessel, B. Normand, F. Mila, and A. Honecker,
Phys. Rev. Res. \textbf{1}, 033038 (2019).

\end{thebibliography}
\end{document}